\documentclass[iop]{emulateapj}                               
\usepackage{ amsmath }
\usepackage{ amssymb }
\usepackage{graphicx}
\usepackage{aas_macros}
\usepackage{subfigure}
\usepackage{morefloats}
\shorttitle{Kompaneets fits to Ori.-Eri. Superbubble II}

\begin{document}
\title{Kompaneets Model Fitting of the Orion--Eridanus Superbubble II: Thinking Outside of Barnard's Loop}
\author{Andy Pon\altaffilmark{1}, Bram B.\ Ochsendorf\altaffilmark{2}, Jo{\~a}o Alves\altaffilmark{3}, John Bally\altaffilmark{4}, Shantanu Basu\altaffilmark{1}, \& Alexander G.\ G.\ M.\ Tielens\altaffilmark{5}}

\altaffiltext{1}{Department of Physics and Astronomy, The University of Western Ontario, 1151 Richmond Street, London, N6A 3K7, Canada; apon@uwo.ca}
\altaffiltext{2}{Department of Physics and Astronomy, The Johns Hopkins University, 3400 North Charles Street, Baltimore, MD 21218, USA}
\altaffiltext{3}{Department of Astrophysics, University of Vienna, T\"{u}rkenschanzstrasse 17, A-1180 Vienna, Austria}
\altaffiltext{4}{Department of Astrophysical and Planetary Sciences, University of Colorado, UCB 389 CASA, Boulder, CO 80389-0389, USA}
\altaffiltext{5}{Leiden Observatory, Leiden University, P.O.\ Box 9513, NL-2300 RA, The Netherlands}

\begin{abstract}
The Orion star-forming region is the nearest active high-mass star-forming region and has created a large superbubble, the Orion--Eridanus superbubble. Recent work by Ochsendorf et al.\ has extended the accepted boundary of the superbubble. We fit Kompaneets models of superbubbles expanding in exponential atmospheres to the new, larger shape of the Orion--Eridanus superbubble. We find that this larger morphology of the superbubble is consistent with the evolution of the superbubble being primarily controlled by expansion into the exponential Galactic disk ISM if the superbubble is oriented with the Eridanus side farther from the Sun than the Orion side. Unlike previous Kompaneets model fits that required abnormally small scale heights for the Galactic disk ($<$40 pc), we find morphologically consistent models with scale heights of 80 pc, similar to that expected for the Galactic disk. 
\end{abstract}


\section{INTRODUCTION}
\label{introduction}

O- and B-type stars create strong stellar winds, intense radiation fields, and powerful supernova explosions. All three of these processes can shape the natal molecular clouds surrounding these stars, with such young stars often creating large cavities of hot (10$^6$ K) plasma in the ISM (e.g. \citealt{Heiles76, McCray87, StaveleySmith97, Heyer98, Churchwell06, Churchwell07, Bagetakos11}). The combined action of an OB association can lead to such bubbles becoming hundreds of parsecs in size. These large bubbles formed by OB associations are referred to as superbubbles.

\begin{figure}[htbp]
   \centering
   \includegraphics[width=3.5in]{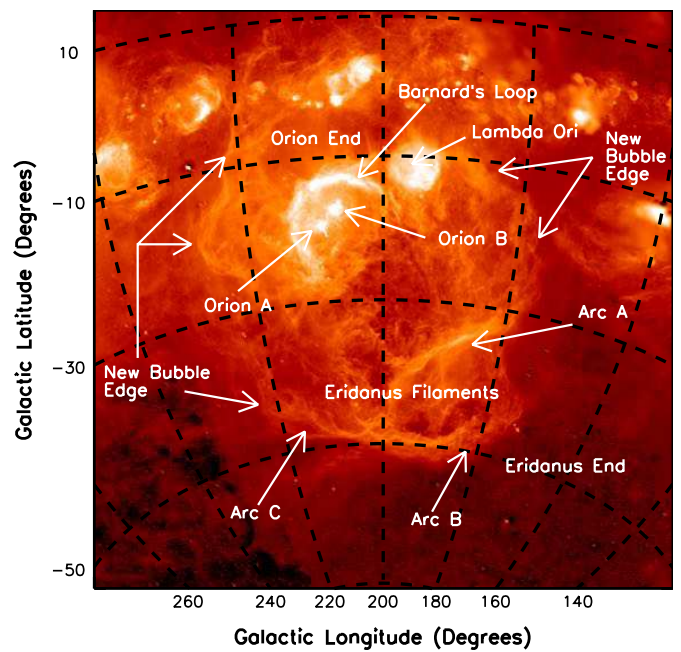}
   \caption{H$\alpha$ map of the Orion--Eridanus superbubble, based on data from the Virginia Tech Spectra Line Survey (VTSS), Southern H-Alpha Sky Survey Atlas (SHASSA) and the Wisconsin H$\alpha$ Mapper (WHAM), as obtained via the Sky View Virtual Observatory. The H$\alpha$ intensities are logarithmically scaled and capped at 200 Rayleighs to highlight the weaker, diffuse H$\alpha$ features coming from the bubble wall. The labels show key features of the superbubble, with the approximate edge of the superbubble proposed by \citet{Ochsendorf15} labeled as the ``new bubble edge.''}
   \label{fig:overview}
\end{figure}

While the Perseus OB2 (300 pc distant) and Sco-Cen (150 pc distant) star-forming regions contain massive stars, have the potential to form additional massive stars, and have formed superbubbles, they are currently only creating low- and intermediate-mass stars \citep{Bally08Walawender, Preibisch08}. The nearest active, high-mass star-forming region is the Orion star-forming region, located approximately 400 pc away from the Sun \citep{Hirota07, Menten07,Sandstrom07}. The Orion star-forming region has created a large superbubble, extending at least 45 degrees into the constellation of Eridanus, such that the superbubble has been named the Orion--Eridanus superbubble (e.g., \citealt{Reynolds79}). Figure \ref{fig:overview} shows a labeled H$\alpha$ image of the Orion--Eridanus superbubble, based on data from the Virginia Tech Spectra Line Survey, Southern H-Alpha Sky Survey Atlas, and the Wisconsin H$\alpha$ Mapper \citep{Dennison98, Gaustad01, Finkbeiner03, Haffner03}. 

In the half of the superbubble closer to the Galactic plane, the Orion side of the bubble, there exists a bright crescent of H$\alpha$ emission that is known as Barnard's Loop \citep{Pickering1890, Barnard1894}, as well as the $\lambda$ Ori ring, a spherical supernova remnant (SNR; \citealt{Morgan55}). Barnard's Loop was previously believed to be the outer wall of the Orion--Eridanus superbubble, with $\lambda$ Ori lying outside of the bubble (e.g., \citealt{Pon14komp}), but \citet{Ochsendorf15} recently suggested that both Barnard's Loop and the $\lambda$ Ori SNR are individual supernova remnants embedded within the larger Orion--Eridanus superbubble. \citet{Ochsendorf15} identified additional H$\alpha$ features between Barnard's Loop and the Galactic plane as the possible edge of the superbubble, rather than these features just being unassociated gas structures illuminated by ionizing photons escaping from within the superbubble. These features are identified in Figure \ref{fig:overview}.

In the side of the superbubble farther from the Galactic plane, the Eridanus side of the bubble, there are three filamentary features referred to as the Eridanus filaments and individually denoted as Arcs A, B, and C \citep{Meaburn65, Meaburn67, Johnson78, Pon14fil}. These are also labeled in Figure \ref{fig:overview}.

In this paper, we readdress the superbubble model solutions found by \citet{Pon14komp} to see if any reasonable fits can be found to the larger superbubble extent suggested by \citet{Ochsendorf15}. In particular, we test whether such a larger superbubble shape reduces the previously noted discrepancy between the required scale height of the ISM in the superbubble models and the generally accepted value \citep{Pon14komp}. 

\section{KOMPANEETS MODEL FITTING}
\label{Kompaneets}

\subsection{Model Set-up}
\label{setup}

The current, standard, analytic model for superbubble growth is the Kompaneets model \citep{Kompaneets60, Basu99}. This model assumes, among other things, that a bubble expands into an exponential atmosphere, that the driving source is stationary with respect to the exponential atmosphere, and that the pressure within the bubble is spatially uniform.

The H$\alpha$ data set presented in Figure \ref{fig:overview} is used for the Kompaneets model fitting in this paper. All fitting is done by eye and we caution that there is significant degeneracy in the models that provide reasonable fits to the H$\alpha$ morphology of the superbubble. 

For model fits presented in this paper, the driving source is required to be located near the Orion A and Orion B molecular clouds, toward the heart of the Orion star-forming region. The Orion end of the superbubble is set at a distance of 400 pc. For the Orion half of the superbubble, the approximate boundaries of the superbubble identified by \citet{Ochsendorf15} are used. That is, the H$\alpha$ features outside of Barnard's Loop are used as the bubble edge, such that the superbubble is wider than Barnard's Loop and wider than in the \citet{Pon14komp} models.

Due to the greater extent of the superbubble in \citet{Ochsendorf15}, Arc C can now be incorporated into the edge of the superbubble wall, as part of a continuous structure with Arc B.  Such a large bubble extent means that the diffuse 0.25 keV X-ray emission located near Arc C can also be encompassed within the bubble boundary \citep{Snowden97}. Previously, \citet{Pon14komp} were forced to place Arc C outside of the superbubble wall, with the suggestion that Arc C was formed from a localized blowout of the superbubble.

While Arc A lies along the wall of the superbubble in these model fits, we do not require it to lie along the edge of the bubble, as seen from the Sun, or trace a surface of equal distance from the driving source, as required in the \citet{Pon14komp} models. There are indeed indications that Arc A may not even be associated with the superbubble \citep{Boumis01, Welsh05, Ryu06, Pon14fil}. 

As in \citet{Pon14komp}, different orientations of the superbubble, with the Eridanus end of the bubble being further or closer to the Sun than the Orion end, are examined. The orientation of the superbubble is parametrized by the inclination, relative to the plane of the sky, of the superbubble's major axis at the point in the sky where the middle of the superbubble would be if the ends were equidistant. This inclination is denoted as $\theta$, such that
\begin{equation}
i = \theta - \sin^{-1}\left(\frac{z}{2 d_{e}}\right),
\end{equation}
where $i$ is the angle between the superbubble's major axis and the plane of the sky at the Orion end of the superbubble (with positive values indicating the bubble is going into the plane of the sky), $z$ is the major axis length of the superbubble, and $d_{e}$ is the distance to the Orion end of the superbubble (400 pc). For this parametrization, negative values of $\theta$ place the Eridanus end closer than the Orion end, positive values place the Eridanus end farther than the Orion end, and for equidistant ends, $\theta$ is 0. We will later introduce an additional angle, $\phi$, to denote the angle between the normal to the Galactic plane and the major axis of the superbubble. Please see \citet{Pon14komp} for more details about the Orion--Eridanus superbubble and the motivations behind fitting a Kompaneets model to the superbubble. 

\subsection{Best Fits}
\label{fits}
\begin{figure*}
\centering
\subfigure[]{
  \includegraphics[width=\columnwidth]{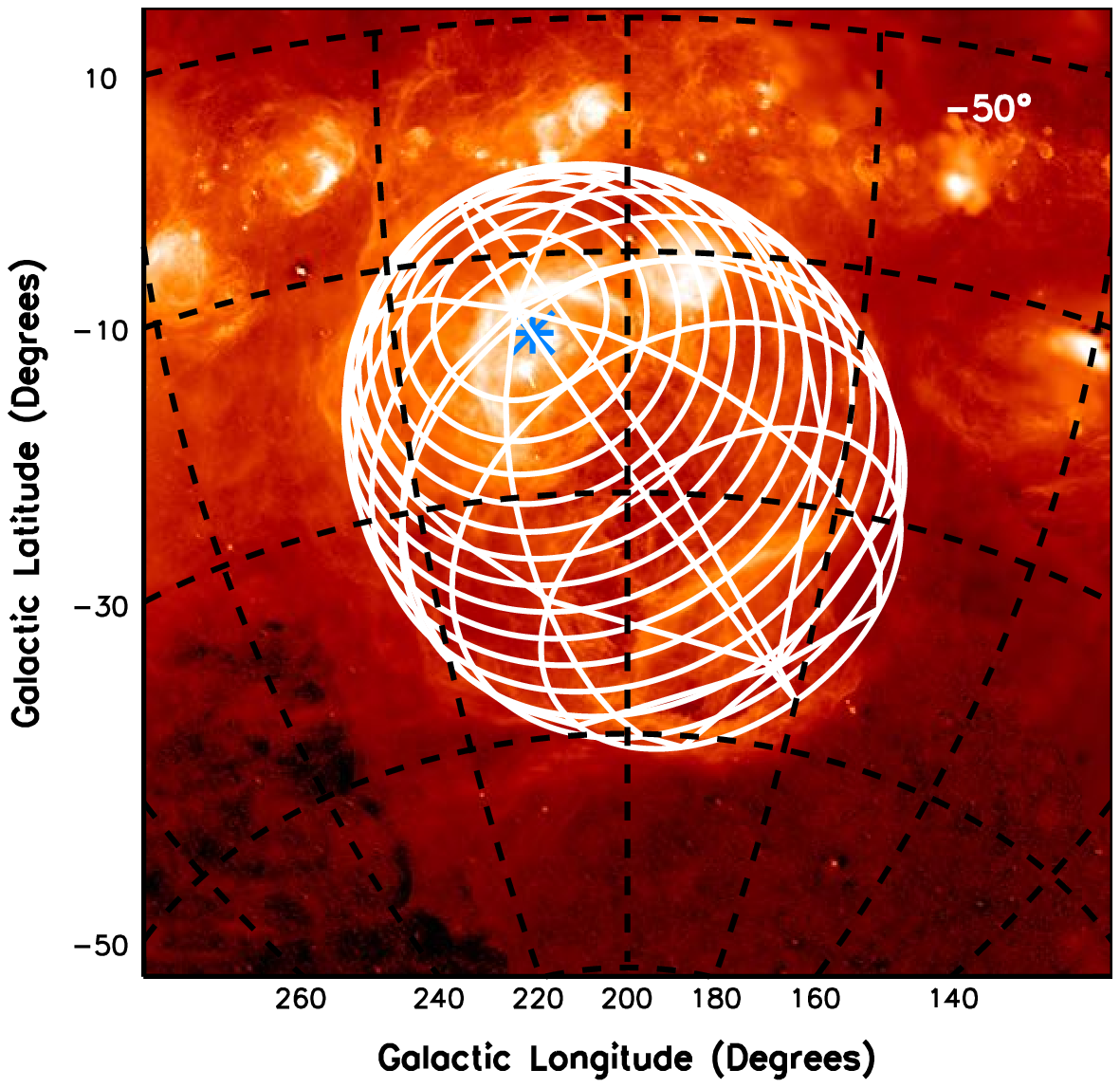}
}
\subfigure[]{
  \includegraphics[width=\columnwidth]{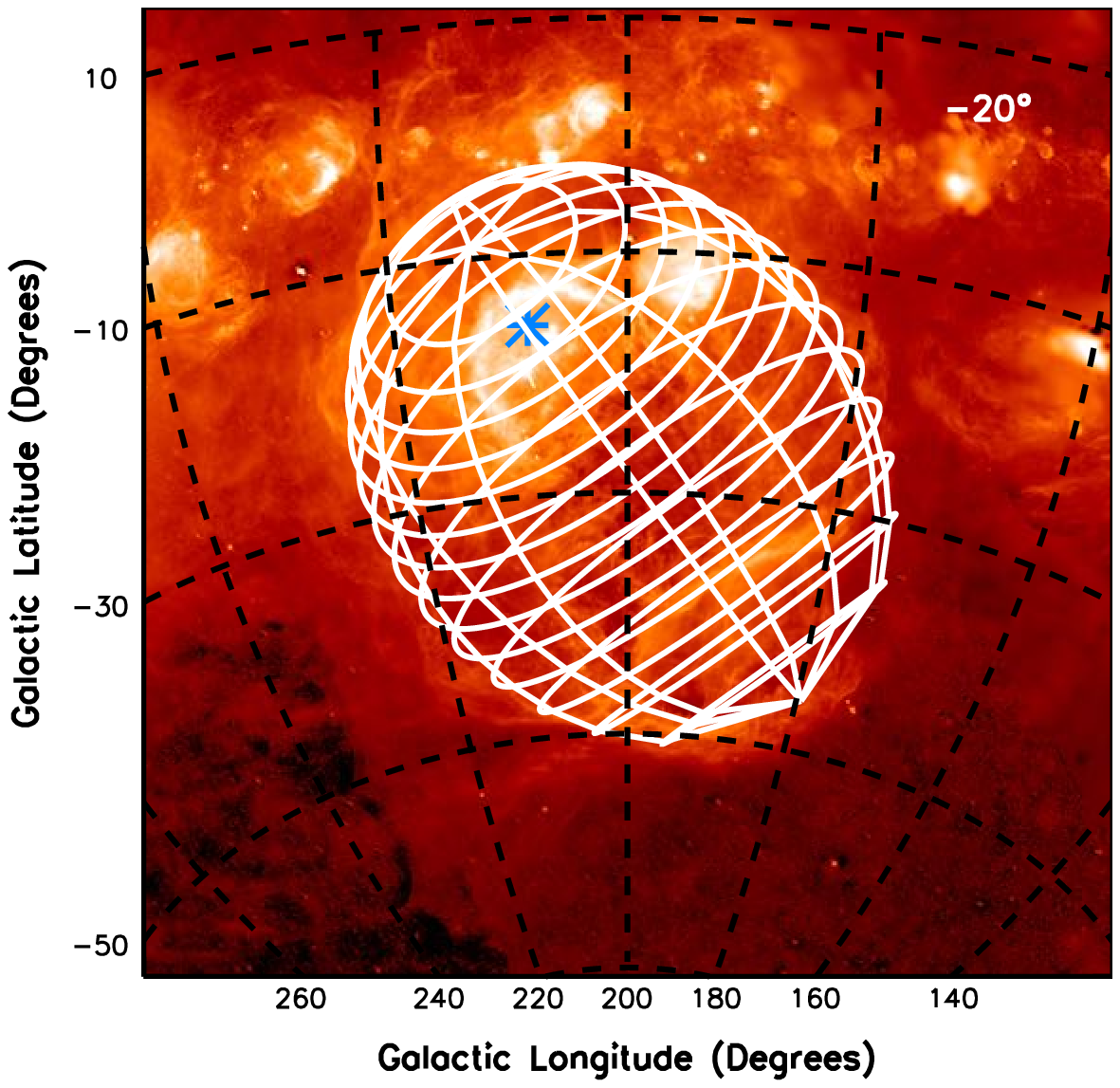}
}
\subfigure[]{
  \includegraphics[width=\columnwidth]{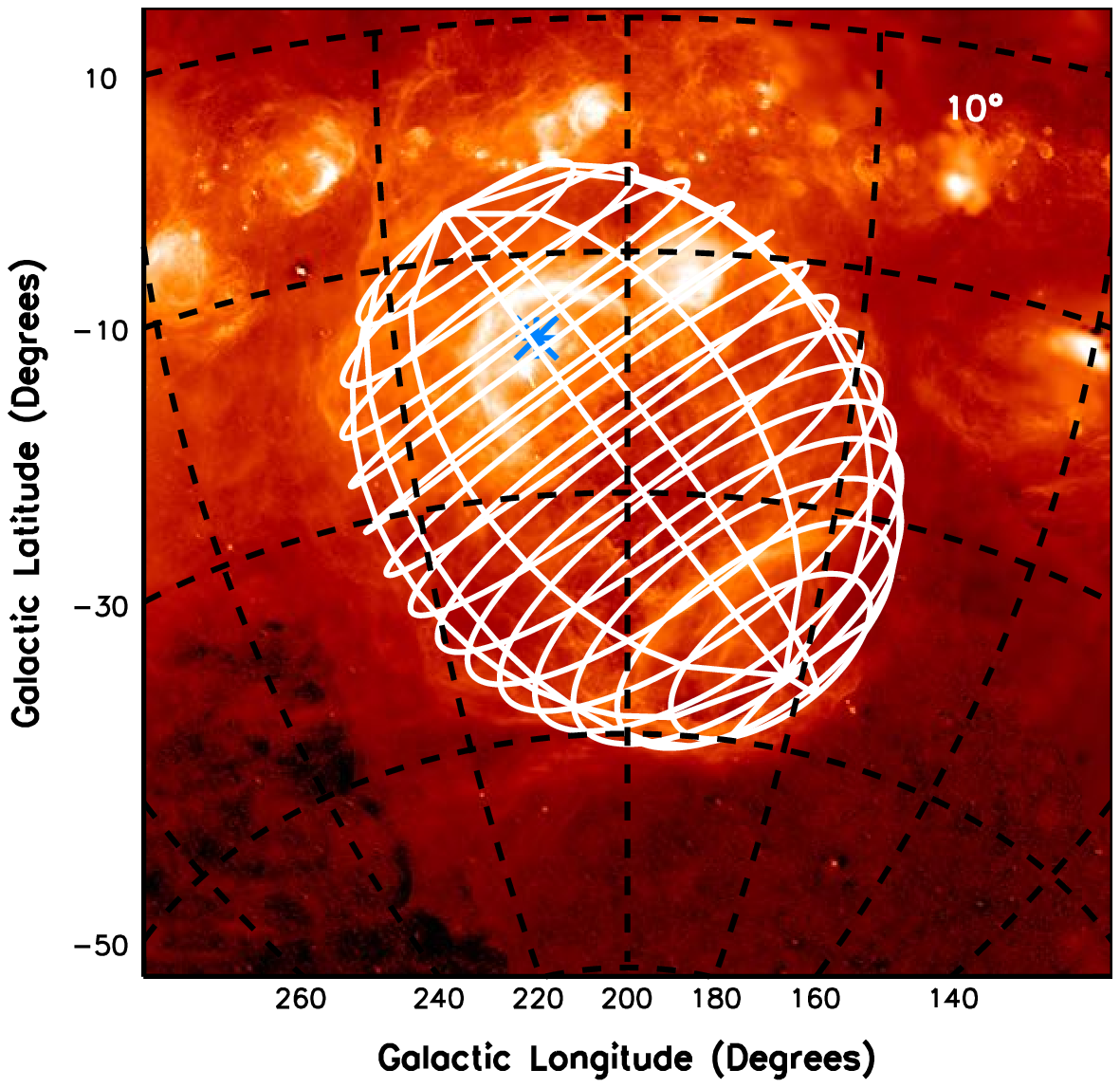}
}
\subfigure[]{
  \includegraphics[width=\columnwidth]{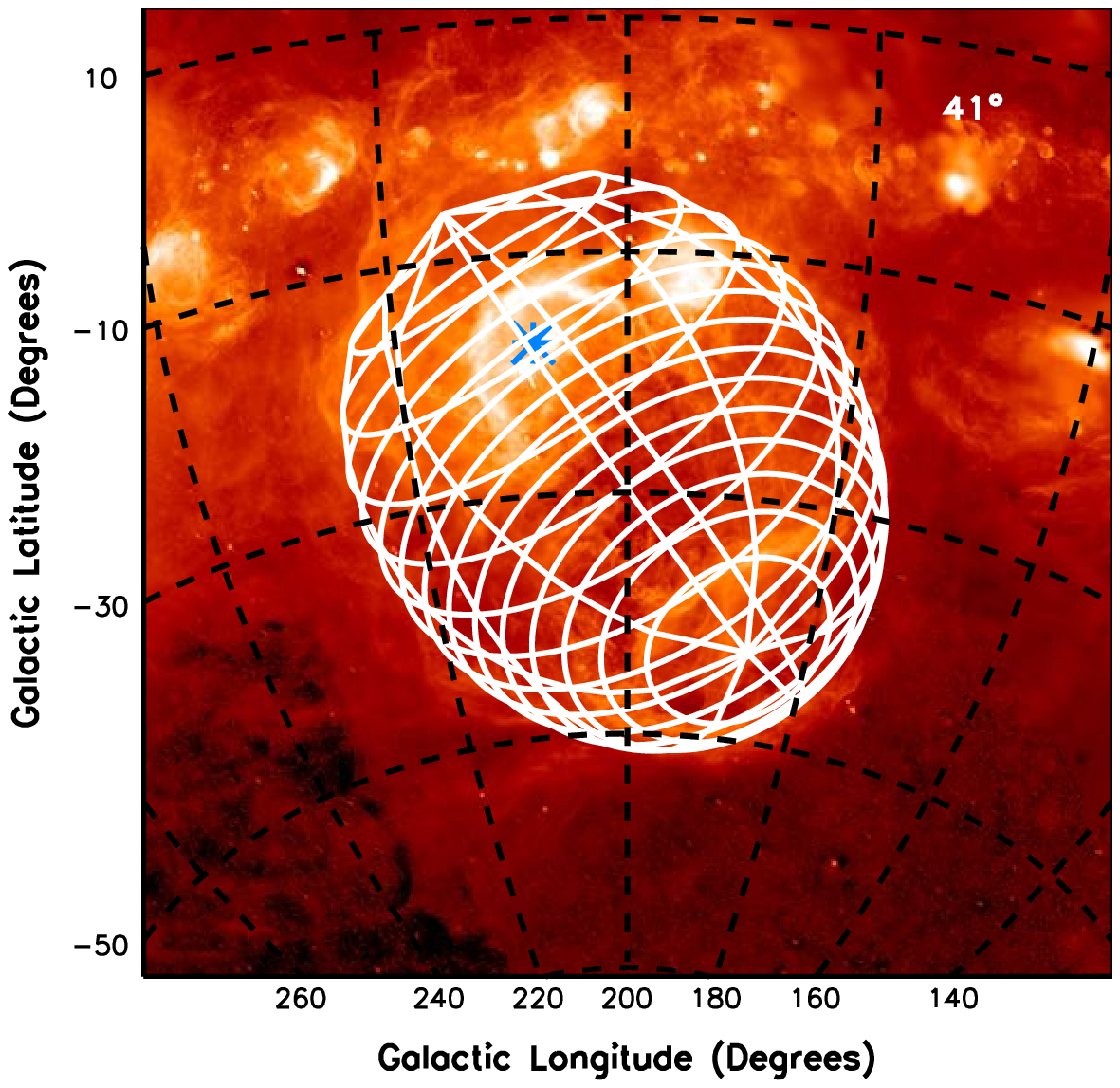}
}
\caption{Best-fit Kompaneets models for $\theta$ = -50$^\circ$, -20$^\circ$, 10$^\circ$, and 41$^\circ$ are shown in panels (a) - (d). The value of $\theta$ is indicated in the top right of each panel. The blue asterisk shows the location of the driving source. The background color is the H$\alpha$ integrated intensity from Figure \ref{fig:overview}.}
\label{fig:models}
\end{figure*}

Reasonable fits to the superbubble morphology are found for most inclinations and a range of the best fitting models are shown in Figure \ref{fig:models}. A summary of the input parameters of the different best fits are listed in Table \ref{table:input}, while a summary of derived parameters are given in Table \ref{table:derived}. 

\begin{deluxetable*}{cccccccc}
\tablecolumns{8}
\tablecaption{Kompaneets model input properties \label{table:input}}
\tablewidth{0pt}
\tablehead{
\colhead{$\theta$} & \colhead{$\tilde y$} & \colhead{$l_{s}$} & \colhead{$b_{s}$} & \colhead{$d_{s}$} & \colhead{$l_{e}$} & \colhead{$b_{e}$} & \colhead{$d_{e}$} \\
\colhead{($^{\circ}$)} & \colhead{} & \colhead{($^{\circ}$)} & \colhead{($^{\circ}$)} & \colhead{(kpc)} & \colhead{($^{\circ}$)} & \colhead{($^{\circ}$)} & \colhead{(kpc)}  \\
\colhead{(1)} & \colhead{(2)} & \colhead{(3)} & \colhead{(4)} & \colhead{(5)} & \colhead{(6)} & \colhead{(7)} & \colhead{(8)}
}
\startdata
-50 & 1.94 & 208.1 & -16.5 & 356 & 209.5 & -14.5 & 400 \\
-35 & 1.87 & 208.3 & -15.3 & 356 & 211 & -11 & 400 \\
-20 & 1.75 & 208.5 & -15.9 & 359 & 213 & -9 & 400 \\
-5 & 1.66 & 207.9 & -16.3 & 369 & 214 & -7 & 400 \\
10 & 1.64 & 207.8 & -17.0 & 385 & 215 & -6 & 400 \\
25 & 1.67 & 207.3 & -16.8 & 405 & 215 & -5 & 400 \\
41 & 1.80 & 208.2 & -17.4 & 423 & 215 & -6 & 400
\enddata
\tablecomments{Column 1 gives the inclination, relative to the plane of the sky, of the superbubble's major axis at the point in the sky where the middle of the superbubble would be if the ends were equidistant, with negative values indicating that the Eridanus end is closer than the Orion end. Column 2 gives the value of the $\tilde y$ parameter. Columns 3-5 give the Galactic longitude, Galactic latitude, and distance from the Sun of the driving source of the superbubble, while Columns 6-8 give the Galactic longitude, Galactic latitude, and distance from the Sun of the Orion end of the bubble.}
\end{deluxetable*}

\begin{deluxetable*}{cccccccccccc}
\tablecolumns{12}
\tablecaption{Kompaneets model derived properties \label{table:derived}}
\tablewidth{0pt}
\tablehead{
\colhead{$\theta$} & \colhead{$H$} & \colhead{$d_\text{min}$} & \colhead{$d_\text{max}$} & \colhead{$\phi$} & \colhead{Age} & \colhead{$P/k$} & \colhead{$T$} & \colhead{$t_\text{blowout}$} & \colhead{$t_\text{super}$} & \colhead{$E_\text{tot}$} & \colhead{$v_\text{exp}$} \\
\colhead{($^{\circ}$)} & \colhead{(pc)} & \colhead{(pc)} & \colhead{(pc)} & \colhead{($^\circ$)} & \colhead{(Myr)} & \colhead{(10$^4$ cm$^{-3}$ K)} & \colhead{(10$^6$ K)} & \colhead{(Myr)} & \colhead{(Myr)} & \colhead{($10^{51}$ erg)} & \colhead{(km s$^{-1}$)}\\
\colhead{(1)} & \colhead{(2)} & \colhead{(3)} & \colhead{(4)} & \colhead{(5)} & \colhead{(6)} & \colhead{(7)} & \colhead{(8)} & \colhead{(9)} & \colhead{(10)} & \colhead{(11)} & \colhead{(12)} 
}
\startdata
-50 & 35 & 134 & 401 & 89 & 2.0 & 1.2 & 3.9 & 0.2 & -0.2 & 1.2 & 205 \\
-35 & 41 & 179 & 417 & 71 & 2.3 & 1.5 & 3.8 & 0.5 & 0.006 & 1.5 & 200 \\
-20 & 54 & 204 & 444 & 58 & 3.2 & 1.6 & 3.6 & 1.3 & 0.6 & 2.0 & 107 \\
-5 & 66 & 224 & 484 & 47 & 4.0 & 1.5 & 3.5 & 2.3 & 1.5 & 2.5 & 77 \\
10 & 75 & 245 & 536 & 38 & 4.8 & 1.3 & 3.4 & 3.0 & 2.0 & 3.0 & 68 \\
25 & 81 & 268 & 599 & 33 & 5.7 & 1.1 & 3.3 & 3.2 & 2.1 & 3.6 & 68 \\
41 & 76 & 294 & 674 & 30 & 5.9 & 0.8 & 3.3 & 2.0 & 1.0 & 3.8 & 98 
\enddata
\tablecomments{Column 1 gives the inclination, relative to the plane of the sky, of the superbubble's major axis at the point in the sky where the middle of the superbubble would be if the ends were equidistant, with negative values indicating that the Eridanus end is closer than the Orion end. Column 3 gives the scale height of the exponential atmosphere. Columns 4 and 5, respectively, give the distance to the closest and farthest point on the superbubble wall from the Sun. Column 6 gives the angle that the major axis of the superbubble makes with the normal to the Galactic plane. Columns 7-9 give the age, interior pressure, and interior temperature of the bubble, respectively, assuming an initial density of the exponential atmosphere of 0.75 cm$^{-3}$ at the height of the driving source and a wind luminosity of $2 \times 10^{37}$ erg s$^{-1}$. Columns 10 and 11 give the time until the superbubble blows out and the time until the top cap becomes supersonic. These times are measured from the present, such that negative values for Column 11 indicate that the superbubble top cap is already supersonic. The total amount of mechanical energy injected into the superbubble so far is given in Column 12. Column 13 gives the model predicted current expansion speed of the Eridanus end of the superbubble.}
\end{deluxetable*}

One significant difference between the various models is the minimum distance between the Sun and the near side of the superbubble. As the Eridanus end of the bubble is moved farther from the Sun, the near side of the superbubble also moves further away. Based upon interestellar absorption features toward the Eridanus half of the superbubble, the near side of the superbubble is typically taken to be at a distance of 180 pc from the Sun \citep{Frisch90,Guo95, Burrows96,Welsh05}. For the near side of the bubble wall to be located 180 pc away, a $\theta$ value of approximately -35$^{\circ}$ is required. Smaller $\theta$ values place the near side too close to the Sun, for instance the best fit with $\theta = -50^{\circ}$ places the near wall at a distance of only 134 pc, while larger values of $\theta$ produce near side distances greater than 200 pc. For the large $\theta$ models, the material causing the absorption feature at 180 pc would have to be a structure separate from the superbubble wall, perhaps associated with the foreground population of low \citep{Bouy14} and high-mass stars \citep{Bouy15}.

The far side of the superbubble, toward the Eridanus filaments, has not been detected in absorption line studies and may reside greater than 500 pc from the Sun  \citep{Boumis01, Welsh05, Ryu06}. Only the models with positive $\theta$ values have maximum distances of the bubble wall from the Sun greater than 500 pc. The negative $\theta$ values would require the backside of the bubble to have been missed in these absorption line studies, potentially due to the wall being too highly ionized to be detected in Na{\sc i} or Ca{\sc ii} absorption. 

The model with $\theta = 41^{\circ}$ is of particular note as this model has the superbubble aligned as close to the normal to the Galactic plane as possible. As discussed in greater detail in \citet{Pon14komp}, because the Orion star-forming region is 130 pc below the Galactic plane, the elongation of the superbubble to more negative Galactic latitudes can either be due to a physical extent perpendicular to the plane or elongation toward the Sun. The $\theta = 41^{\circ}$ model minimizes the angle between the superbubble major axis and the normal to the Galactic plane, although this angle is constrained to be at least 30$^{\circ}$ based on the angle between the projection of the superbubble major axis on the plane of the sky and the Galactic normal. Models with smaller values of $\theta$ are more closely aligned parallel to the Galactic plane, with the $\theta = -35^{\circ}$ model making an angle of 71$^{\circ}$ with respect to the Galactic normal. The $\theta = -50^{\circ}$ model has the superbubble almost perfectly parallel to the Galactic plane.

The parameter $\tilde y$ is a measure of the relative evolutionary stage of a superbubble, with $\tilde y$ increasing from 0 to 2, at which point the bubble has completely blown out. When coupled with the maximum physical radius of a superbubble, $R$, the $\tilde y$ parameter can be used to determine the required scale height $H$ of the exponential atmosphere into which the superbubble is expanding, via
\begin{equation}
R = 2 H \sin^{-1}\left(\tilde y\right).
\end{equation}
Prior observations of the ISM in the Milky Way suggest that the Galactic ISM should have a scale height of the order of 100 to 150 pc in the vicinity of Orion \citep{Kalberla09}. 

For the models presented in this paper, the elongation of the superbubble is the smallest when the absolute value of $\theta$ is small, with $\tilde y$ correspondingly increasing with the absolute magnitude of $\theta$. For the models with $\theta$ between -20$^\circ$ and $-50^{\circ}$, the required scale height of the ISM is a problematically small 35-55 pc. For positive values of $\theta$, the increase in $\tilde y$ with $|\theta|$ is partially offset by the increasing physical radius of the superbubble, as a more distant bubble must be larger to have the same angular size. As such, most of the $\theta > 0^{\circ}$ models have very similar scale heights around 75-80 pc. While not quite the expected 100 pc scale height of the Galaxy, this is much closer to what is expected than required by the previous narrower Kompaneets models of \citet{Pon14komp}, which had scale heights of at most 40 pc.

The dimensionless parameters of a Kompaneets model can be converted to physical units if the initial density of the ISM at the height of the driving source and the mechanical energy input rate are given. The age, interior pressure, and interior temperature of each superbubble model are given in Table \ref{table:derived} under the assumption that the initial gas density is 0.75 cm$^{-3}$ \citep{Heiles76, Ferriere91, Brown95, Kalberla09} and the mechanical energy input rate of the Orion star-forming region is $2 \times 10^{37}$ erg s$^{-1}$ \citep{Reynolds79, Brown94}. This is the mechanical energy input rate and the midpoint of the density range investigated by \citet{Pon14komp}. 

The age of the superbubble increases with increasing $\theta$, with ages from 2.0 to 6.0 Myr. These values are all consistent with previous estimates of a few million years for the dynamical age of the superbubble \citep{Brown94}, as well as the range of ages of the various stellar groups in Orion. The OB1a, b, c, and d groups have ages of 8-12, 2-8, 2-6, and $<2$ Myr, respectively \citep{Brown94, Bally08}. The bubble models predict interior pressures, $P / k$, of the order of 10$^4$ cm$^{-3}$ K and interior temperatures of (3-4) $\times 10^{6}$ K, consistent with prior estimates \citep{Williamson74, Naranan76, Long77, Burrows93, Guo95, Burrows96}. 

Table \ref{table:derived} also gives the time until the superbubble models will blow out, the time until the top cap will become supersonic, and the total energy injected into the superbubble. These times are measured from the present epoch, rather than from the birth of the superbubble. For all models, the total energy required to form the bubble is of the order of a few times 10$^{51}$ erg, which can be provided by a small number of supernova explosions, given that a typical supernova injects at most 10$^{51}$ erg (e.g., \citealt{Veilleux05}). This is also of the order of the estimated kinetic energy of the superbubble from observations \citep{Brown95}. 

The fundamental assumption of the Kompaneets model that the interior pressure of the superbubble remains spatially uniform is expected to break down when the expansion velocity becomes larger than the interior sound speed. For the two best-fit models presented in \citet{Pon14komp}, the end cap becomes supersonic before the bubble takes its final, observed morphology. For the models presented in this paper, the expansion speed is subsonic at all times, except for the $\theta = -50^{\circ}$ model. Table \ref{table:derived} gives the predicted expansion velocity of the Eridanus end cap at the current time. For the $\theta = -50^{\circ}$ model, the sound speed is given, instead of the supersonic velocity of the model. The $\theta = -35^{\circ}$ model has an expansion speed approximately equal to the sound speed ($\sim 200$ km s$^{-1}$), while all other models predict lower velocities, in the range of 60-110 km s$^{-1}$. The sound speed is calculated for each bubble based on the model derived interior temperature.

The expansion speed of the superbubble was initially estimated to be 15 km s$^{-1}$ \citep{Menon57, Reynolds79}, but more recent estimates place the expansion velocity closer to 40 km s$^{-1}$ \citep{Cowie78, Cowie79, Brown95, Huang95, Welty02, Ochsendorf15}.  The comparison of these observationally determined expansion velocities and model predictions is hampered by a number of factors. The observed expansion velocities are only based on line-of-sight motions and may be lower than the true, 3D expansion velocities. The brightest emission is also detected toward the edges of the bubble, due to an increase in the line-of-sight depth of the bubble wall, which is where the expansion velocity is preferentially in the plane of the sky. The H$\alpha$ emission from the bubble wall is quite weak toward the interior of the superbubble, making measurements of the expansion velocity difficult for points where the bubble is expanding preferentially along the line of sight. The expansion velocity predicted for the edge of the bubble is the highest at the Eridanus end and decreases toward the Orion end, but most of the measurements of the expansion velocity of the superbubble have been made toward the Orion end. Since all of these effects can lower the observed expansion velocity, it is not clear if the larger expansion velocities predicted from the Kompaneets model are completely at odds with the observations. 

The true expansion velocity of the superbubble, however, is likely to be slightly less than predicted, since the Kompaneets model does not account for momentum conservation. The Kompaneets model also does not account for the cooling of the bubble via mass loading, which would reduce the internal pressure of the bubble and thus reduce the expansion velocity. \citet{Ochsendorf15} present evidence for a lower temperature within the superbubble between Barnard's loop and the Galactic plane and discuss the possibility of a density gradient set up by mass loading within this portion of the superbubble.

\subsection{Ionization Front}
\label{front}
For a particular Kompaneets model, if the temperature of the bubble wall, the ionizing luminosity of the driving source, the pressure within the wall, and the initial density of the surrounding material at the location of the driving source are known, then the locations where the ionizing radiation fully penetrate the bubble wall can be calculated. We adopt a wall temperature of 8000 K \citep{Basu99} and an ionizing luminosity of $4 \times 10^{49}$ s$^{-1}$ \citep{Reynolds79}, and investigate wall pressures in the range of (1-5) $\times 10^4$ K cm$^{-3}$ \citep{Burrows93, Guo95, Burrows96} and initial gas densities between 0.5 and 1 cm$^{-3}$ \citep{Heiles76, Ferriere91, Brown95, Kalberla09}. These are the same ranges used in \citet{Pon14komp}.

For all best-fit models, the ionizing photons will not be fully trapped anywhere within the superbubble if the pressure and density are the minimum values in the above range. As $\theta$ increases, the physical bubble radius also increases, thereby decreasing the ionizing flux at the bubble wall and increasing the surface density of the wall. Models with larger values of $\theta$ can thus more easily trap photons within the bubble wall. For models with $\theta \ge 5^\circ$, the ionizing photons are fully trapped throughout the entirety of the bubble for the largest density and pressure values examined. For these models, since there are reasonable densities and pressures that can lead to the ionizing photons being trapped everywhere or not being fully trapped anywhere, there should be intermediate densities and pressures that will allow the ionizing photons to breakout at any desired point along the superbubble wall.

\citet{Pon14komp} identified linear H {\sc i} features extending radially away from the Orion star-forming region, with significant H {\sc i} emission to more positive Galactic latitudes of these features and very little emission to more negative latitudes. These H {\sc i} features coincide with sharp drops in the H$\alpha$ emission of Barnard's Loop and were interpreted as being due to the ionizing photons breaking out of the Orion--Eridanus superbubble's walls. All of these features, however, now occur within the larger region identified as the Orion--Eridanus superbubble in this paper, such that we no longer have any obvious observational signatures of ionizing photons breaking out of the bubble in order to further constrain the pressure and density of the superbubble model. The total ionizing luminosity of the Orion star-forming region is known to be sufficient to account for the total amount of H$\alpha$ emission observed from the Orion--Eridanus superbubble \citep{Ochsendorf15}.

\section{COMPARISON WITH PREVIOUS MODELS}
\label{comparison}

\citet{Pon14komp} previously fit Kompaneets models to the Orion--Eridanus superbubble, but assumed that Barnard's Loop was part of the superbubble wall. As such, the \citet{Pon14komp} model fits are more elongated, have larger $\tilde y$ values, and have smaller scale heights (15-40 pc) than the best-fit models presented in this paper. 

\citet{Pon14komp} presented two best fitting models: one where the Eridanus side of the superbubble is closer to the Sun than the Orion star-forming region (denoted as model T) and one where the Eridanus side is more distant (denoted as model A). Model T is relatively similar to the $\theta = -50^{\circ}$ and $\theta = -35^{\circ}$ models, as all three models have the superbubble inclined almost parallel to the Galactic plane and have scale heights much less than 100 pc (e.g., 15 pc in the case of model T). Model A is most similar to the $\theta = 10^{\circ}, 25^{\circ}$ and $41^{\circ}$ models, with these four bubbles all being relatively closely aligned to the normal to the Galactic plane and having some of the larger scale heights of the best fitting models. \citet{Pon14komp} found that to get a good fit with the Eridanus side more distant, that is for model A, they required a bubble that is larger than Barnard's Loop at more positive Galactic latitudes. To explain this size mismatch, they suggested that extra material in the Orion star-forming region could have preferentially hindered the expansion of the bubble toward the Galactic plane. The model A fit, however, did not quite extend out to the H$\alpha$ features that we consider to be the edge of the superbubble in this paper. 

In \citet{Pon14komp}, Arcs A and B were required to trace a constant line of latitude along the edge of the superbubble, a constraint not required in this paper. \citet{Pon14komp} thus ascribed the formation of Arcs A and B to a process dependent on the distance from the driving source of the superbubble, such as the breakout of the ionizing flux. This also required one of the two arcs to be on the near side of the superbubble and one to be on the far side. 

In the models presented in this paper, Arc A is not constrained to lie on the near or far sides of the bubble. Therefore, the different models make no prediction about whether Arc A should have the velocity of the near or far side of the superbubble or whether Arc A should absorb X-rays coming from the hot interior of the superbubble. There is some debate about whether Arc A is even associated with the superbubble \citep{Pon14fil}. 

The models presented in this paper do, however, provide a natural explanation for the appearance of Arcs B and C. In all of the models, these arcs lie along the edge of the bubble, where the line of sight through the bubble wall should be lengthened and the bubble most visible. That is, these models suggest that Arcs B and C are visible due to geometric projection affects, rather than the Arcs having to be regions with more mass than their surroundings, as in the \citet{Pon14komp} models.

\begin{figure*}[htbp]
   \centering
   \includegraphics[width=7in]{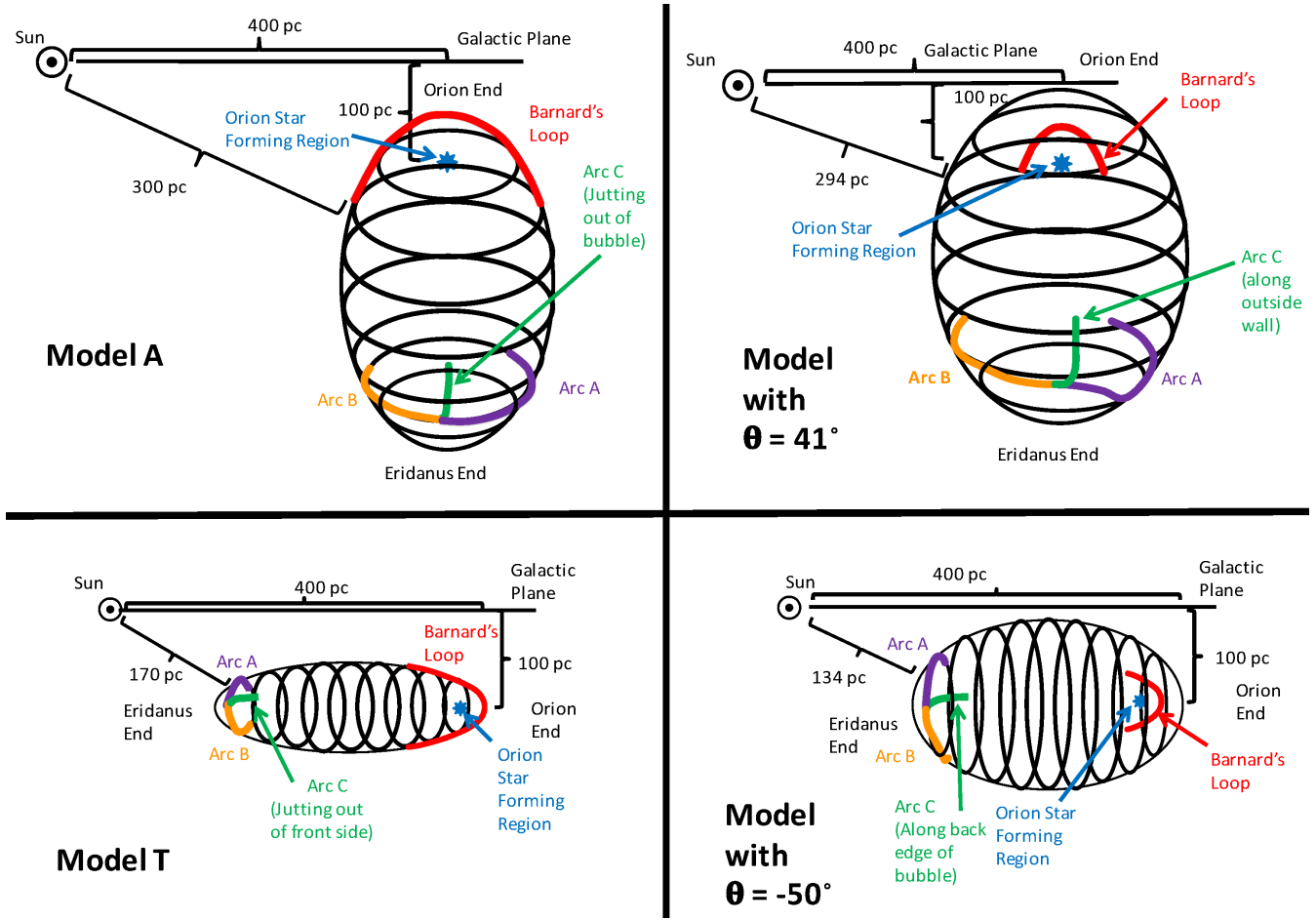}
   \caption{Cartoon diagrams of models A and T from \citet{Pon14komp} and the models with $\theta = 41^{\circ}$ and $-50^{\circ}$ from this paper. Key features of the bubble are identified and labeled. In these diagrams, the bubbles must still be inclined out of the plane of the page, with the Eridanus side further out of the page, to match the roughly 30$^{\circ}$ angle between Galactic normal and the projection of the bubble major axis on the plane of the sky. }
   \label{fig:schematics}
\end{figure*}

Cartoon schematics of the $\theta = -50^{\circ}$ and $41^{\circ}$ models are shown in Figure \ref{fig:schematics}, along with diagrams for models A and T of \citet{Pon14komp}. 

\section{DISCUSSION}
\label{discussion}

The models with the most positive and most negative $\theta$ values present very qualitatively different bubble morphologies for the Orion--Eridanus superbubble. The models with the most positive $\theta$ values have the bubble oriented roughly perpendicular to the Galactic plane and produce scale heights close to, albeit slightly smaller than, the 100 pc scale height expected for the Galactic disk. These model predictions for the structure of the Galactic ISM are consistent with the expected structure of the ISM. Therefore, if the Orion--Eridanus superbubble morphology is that of these positive $\theta$ models, Kompannets model should be considered to be reasonable representations of the superbubble, further meaning that the expansion of the superbubble has likely been primarily controlled by the exponential density gradient of the Galactic disk. A small, additional contribution from a secondary factor, such as magnetic fields \citep{Tomisaka92, Tomisaka98,Stil09}, would still be required to explain the small ($\sim 30^{\circ}$) angle between the normal to the Galactic plane and the superbubble major axis. 

The presence of a series of bubbles nested within the Orion--Eridanus superbubble, such as Barnard's Loop and the $\lambda$ Ori bubble \citep{Ochsendorf15}, is consistent with the Kompaneets model, as the model assumes a constant energy input from the driving source. These additional supernova explosions within the larger superbubble will rapidly expand and transfer energy to the superbubble wall, thereby providing additional energy to the superbubble over the lifetime of the Orion star-forming region. This rejuvenation of the superbubble from successive supernovae is further discussed in \citet{Ochsendorf15}.

The models with the most negative $\theta$ values require a density gradient parallel to the Galactic plane with an unlikely small scale height of $\sim$40 pc, in significant disagreement with the expected density structure in the Galactic plane. As such, if the Orion--Eridanus superbubble is indeed oriented parallel to the plane, we do not consider Kompaneets models to be a good fit to the superbubble. To create such a parallel bubble, a physical mechanism not included in the Kompaneets model is likely primarily controlling the evolution of the superbubble. 

One possible explanation for such an elongated superbubble would be if the driving source of the bubble was moving parallel to the plane. The successive supernova explosions from such a moving source would create a chain of adjacent bubbles that could then merge to produce a superbubble elongated parallel to the plane. \citet{Bouy15}, in fact, have recently identified a blue stream of young stars extending into the plane of the sky toward the Orion star-forming region, suggesting that star formation has indeed propagated from a position closer to the Sun to the current site of the Orion star-forming region. The nested shells seen by \citet{Ochsendorf15} would then just be the most recent SNRs in a series of supernovae extending away from the Sun in the direction of Orion. \citet{Welsh05} also argued for the presence of multiple gas shells within the superbubble based upon absorption line data and the existence of the Eridanus filaments has also previously been interpreted as evidence for multiple bubbles \citep{Boumis01, Welsh05, Ryu06, Ryu08, Jo11}. Such an orientation of the superbubble parallel to the Galactic plane is plausible, as many other H{\sc i} shells show alignment parallel, rather than perpendicular to the Galactic Plane \citep{Heiles79, Ehlerova05, Ehlerova13, Suad14}. 

Please also see the more in depth discussion within \citet{Pon14komp} of possible secondary driving sources and additional physical processes that could elongate a superbubble. 

\section{CONCLUSIONS}
\label{conclusions}

The Orion star-forming region is the nearest star-forming region actively forming high-mass stars. It has created a large superbubble known as the Orion--Eridanus superbubble. Based on the recently proposed larger size of the superbubble \citep{Ochsendorf15}, we have fit Kompaneets models to the H$\alpha$ delineated shape of the superbubble. We find that the Orion--Eridanus superbubble can be matched by a variety of models with various inclinations with respect to the plane of the sky. 

Models with the Eridanus side closer than the Orion star-forming region ($\theta < 0$) are more consistent with absorption measurements indicating that the near side of the superbubble is 180 pc distant from the Sun \citep{Frisch90,Guo95, Burrows96,Welsh05}, but produce bubbles that are roughly parallel to the Galactic plane and that require unusually small scale heights for the Galactic ISM. Such models are not consistent with the assumption that the superbubble's evolution is dominated by pressure driven expansion into the exponential ISM of the Galactic disk, which predicated the use of a Kompaneets model to fit the superbubble. This morphology of the superbubble could instead potentially indicate a moving driving source, related to the production of the blue streams identified by \citet{Bouy15}.

Models in which the Eridanus side is farther away, however, not only place the major axis of the superbubble reasonably close to the normal to the Galactic plane (as close as 30$^\circ$), but also produce scale heights (80 pc) that are reasonably consistent with the known properties of the Galactic ISM. Previous Kompaneets model fits to the Orion--Eridanus superbubble, where smaller H$\alpha$ extents were used, were unable to produce scale heights larger than 40 pc, regardless of the orientation of the superbubble. We thus posit that if the superbubble is aligned with the Eridanus half farther from the Sun than the Orion half, the decrease in ISM density away from the Galactic plane could be primarily responsible for the current morphology of the superbubble. Only a minor secondary process would be required to explain the slightly small 80 pc scale height, compared to the expected 100-150 pc scale height of the ISM, and the 30$^\circ$ tilt of the superbubble major axis on the plane of the sky. For instance, the local scale height near Orion may be somewhat smaller than the Galactic average of 100-150 pc or magnetic fields could have helped channel the superbubble.

\acknowledgements
	A.P.\ would like to acknowledge that partial salary support was provided by a Canadian Institute for Theoretical Astrophysics (CITA) National Fellowship. This research has made use of the Smithsonian Astrophysical Observatory (SAO) / National Aeronautics and Space Administration's (NASA's) Astrophysics Data System (ADS). This research has made use of the astro-ph archive. The Wisconsin H-Alpha Mapper (WHAM) is funded by the National Science Foundation. We would like to thank the anonymous referee for many useful changes to this paper. 
	
\bibliographystyle{apj}
\bibliography{ponbib}

\end{document}